\documentclass[11pt]{article}

\begin{document}
\vskip 20pt

\centerline{\bf \large Bell Inequality and  Many-Worlds Interpretation}
\vskip 30pt

 \centerline{\bf L. Vaidman}
\centerline{ Raymond and Beverly Sackler School of Physics and Astronomy}
 \centerline{Tel-Aviv University, Tel-Aviv 69978, Israel}
\vskip 20pt
It is argued that the lesson we should learn from Bell's inequalities is not that quantum mechanics requires some kind of action at a distance, but that it leads us to believe in parallel worlds.

\section{Introduction}

  Bell's work \cite{Bell64} led to a revolution in our understanding of Nature.
I remember   attending my first physics conference on ``Microphysical Reality and Quantum Formalism,'' in Urbino, 1985. Most of the talks were about  Aspect's experiment \cite{Aspect} confirming the nonlocality of quantum mechanics based on the experimental violations of Bell's inequalities. Although I did not share the skepticism of many speakers regarding the results of Aspect, I was not ready to accept that a local action in one place can instantaneously change anything at another place. So, while for the majority the lesson from Bell was that  quantum mechanics requires some ``spooky action at a distance'', I was led by Bell's result to an alternative revolutionary change in our view of Nature. I saw no other way, but accepting  the many-worlds interpretation (MWI) of quantum mechanics \cite{Eve,SEP}.

I shall start by presenting the Einstein-Podolsky-Rosen (EPR) \cite{EPR} argument. Then, Bell's idea will be presented using the Greenberger-Horne-Zeilinger setup \cite{GHZ} in the form proposed by Mermin \cite{Mermin,myGHZ}.  The discussion of nonlocality will suggest that Bell's inequalities are the only manifestation of action at a distance in Nature. The demonstration of the necessity of action at a distance  will be done through a detailed analysis of the GHZ experiment. Then, I shall show  how multiple worlds resolve the problem of action at a distance. After discussing the issue of nonlocality in the MWI  I shall conclude by  citing Bell's view on the MWI.

\section{EPR - Bell - GHZ }\label{E-B-G}

The story of Bell cannot be told without first describing the EPR argument. Instead of following the historical route,   I shall use the GHZ setup, which, in my view, is the clearest way to explain the EPR and Bell's discovery.

There are three separate sites with Alice, Bob and Charley which share an entangled state of three spin-$\frac{1}{2}$ particles, the GHZ state:
\begin{equation}
\label{GHZ}
|GHZ\rangle = {1\over \sqrt
  2}{\Large (}|{\uparrow}_z\rangle_A|{\uparrow}_z\rangle_B|{\uparrow}_z\rangle_C -
|{\downarrow}_z\rangle_A|{\downarrow}_z\rangle_B|{\downarrow}_z\rangle_C{\Large
)} .
\end{equation}
The GHZ state is a maximally entangled state of a spin at every site with spins in the two other sites. Therefore, the measurement of the spin in each site and in any direction can be performed, in principle, using measurements at other sites. The assumption that there is no action at a distance in Nature, tells us that the measurements of Alice and Bob cannot change Charley's spin. After Alice's and Bob's measurements,  Charley's spin becomes known. The spin value could not have been changed by distant measurements, therefore it existed before. This is the consequence of the celebrated EPR criterion for a physical reality. According to the EPR argument, the values of the spins of Alice, Bob and Charley in all directions  are elements of reality. Quantum mechanics does not provide these values. Furthermore, the uncertainty relations prevent the simultaneous existence of some of these spin values. Thus, EPR concluded that quantum theory is incomplete.

 At the end of their paper, EPR expressed hope that one day quantum theory will be completed to make these elements of reality certain. It took almost thirty years before Bell showed that it cannot be done.

 We need not  consider many elements of reality to show the inconsistency. In the GHZ setup it is enough to consider the spin values  just in two directions, $x$ and $y$. Let us rewrite the GHZ state in the $x$ basis:
\begin{equation}
\nonumber
 {1\over
  2}(|{\uparrow}_x\rangle_A
|{\uparrow}_x\rangle_B|{\downarrow}_x\rangle_C
 + |{\uparrow}_x\rangle_A
|{\downarrow}_x\rangle_B|{\uparrow}_x\rangle_C
 +
|{\downarrow}_x\rangle_A
|{\uparrow}_x\rangle_B|{\uparrow}_x\rangle_C +
|{\downarrow}_x\rangle_A
|{\downarrow}_x\rangle_B|{\downarrow}_x\rangle_C).
\end{equation}
We see that the product of the spins  measured in the $x$ direction is $-1$ with certainty.
Similarly, if we use the $x$ basis for Alice and $y$ bases for Bob and Charley, we learn that the product of the spins measured in the $x$ direction for Alice, and in the $y$ direction for Bob and Charley, is $1$ with certainty. Due to symmetry of the GHZ state, the product is $1$ with certainty also if it is Bob or Charley, instead of Alice, who  is the only one to measured the spin in $x$ direction. Therefore, the following equations should be fulfilled for the results of the spin measurements:
\begin{eqnarray}
 \{{\sigma_A}_x\} \{{\sigma_B}_x\} \{{\sigma_C}_x\} & = &  -1 , \label{E1}\\
 \{{\sigma_A}_x\} \{{\sigma_B}_y\} \{{\sigma_C}_y\} & = & 1 ,\label{E2}\\
\{{\sigma_A}_y\} \{{\sigma_B}_x\} \{{\sigma_C}_y\} & = & 1 , \label{E3}\\
 \{{\sigma_A}_y\} \{{\sigma_B}_y\} \{{\sigma_C}_x\} & = & 1 ,\label{E4}
\end{eqnarray}
\noindent
where $\{{\sigma_A}_x\}$ signifies the outcome of the measurement  of
$\sigma_x$ by  Alice, etc.
 All values of  spin components  in the above equations are  EPR elements of reality. The outcomes should exist prior to the measurement and independent of what is done to other particles. So, according to EPR, the value of Alice's spin $\{{\sigma_A}_x\}$ appearing in (\ref{E1}) should be the same as in equation (\ref{E2}) and similarly for other values of spin variables. But this contradicts the fact that equations (\ref{E1}-\ref{E4}) cannot be jointly satisfied: the product of the lefthand sides is a product of squares, so it is positive, while the product of the righthand sides is $-1$.

\section{From Bell inequality to nonlocality }

Apparently, the first  conclusion which can be reached here is that Nature is random.

The predictions of quantum theory including   the results (\ref{E1}-\ref{E4}) were tested and verified in all experiments performed to date. We have shown above that equations (\ref{E1}-\ref{E4}) are inconsistent with the assumption that there are definite predictions for all these results. Therefore, at least some of the results should not exist prior to the measurement: the outcome of the measurement is random!

However, there is a problem with this ``proof'' of randomness. Let us assume that Charley's outcome is the one which is random. This  contradicts the fact that after Alice's and Bob's measurements, Charley's result is definite. A nonlocal action is then required to  fulfill the equation. But if nonlocality is accepted, the EPR concept of elements of reality loses its basis, so the proof of randomness fails.

This is a proof of nonlocality. There is no way to explain these quantum correlations by some underlying local definite values or local probability distributions. It seems that the conclusion must be that actions (measurements in $x$ or $y$ directions) of Alice and Bob change the outcome of Charley's measurement performed immediately after. It is a proof that there is an action at a distance. It got the name of ``spooky action at a distance'' since there is no known underlying mechanism, and furthermore,  since it is not observable: one cannot send signals to Charley using measurements in Alice's and Bob's sites.

\section{Against nonlocality }

The role of science is to explain    how  and why things happen. A bird falls because a bullet hits it. The hunter shot the bullet. He was able to point his gun since photons reflected by the bird reached his eyes. In all these explanations, objects are present in particular places and they  interact with other objects by sending particles (photons, bullets) from one object to another. This  allows the concept of location of an object: it is the place where it can be influenced directly by  other objects and where it can influence directly other objects. Clearly, this is the picture in classical physics: only local actions exist.

It is true that classical physics has also global formulations. A minimal action principle provides a complete solution given initial conditions without presenting explicit local mechanism.  There is a logical option for existence of a world  described by an action principle with nonlocal interactions as well as a world with local interactions but without minimal action principle. I feel that the local action explanation is the most important part of our picture of Nature and thus  we should try to keep it even when we turn to the correct physical theory which is not classical, but quantum.

In quantum mechanics,  the Aharonov-Bohm effect \cite{AB} (AB) seems to be a counter example. An electron changes its motion as a function of the magnetic field in a region where the electron does not pass. There is a nonzero vector potential at the location of the electron, but this potential  is not locally defined, only the line integral of the potential is physical (gauge invariant). Recently, I solved, at least for myself, this apparent conflict with locality. I found local explanations of both the scalar and the magnetic  AB effects \cite{VAB}. The electron moves in a free-field region, but the source of the potential, the solenoid, or the capacitor, feels the field of the electron. I considered the latter as  quantum objects and realized that there is an unavoidable entanglement between them and the electron during the AB interference experiment. I calculated the phase acquired by the solenoid in the AB experiment and found that it equals exactly to the AB phase. In the AB experiment the phase is manifested in the interference shift of the electron, but it is explained by the local action of the electromagnetic field of the electron on the solenoid.

Thus, the only manifestation of a nonlocal action  that we know of in physics, is the unexplained nonlolcal Bell-type correlations. Since physics has no mechanism for nonlocal actions, its existence is appears to be a ``miracle''. In other words, physics cannot explain it. As a physicist, I want to believe that we do understand Nature. Bell apparently tells us that it is impossible, or that we need to make a large conceptual change in our views of Nature. The best option I see in this situation is to reject a tacit assumption, necessary for the Bell's proof, that there is only one world. There is one physical Universe, but there are many outcomes in every quantum experiments, corresponding to many worlds as we perceive them.

\section{From Bell inequalities to the MWI }

How exactly the MWI resolves the difficulty due to the EPR-Bell-GHZ argument? The concept of the EPR element of reality is the core of the breakdown.
\begin{quote}
{\it   If, without in any way disturbing a system, we can predict with certainty (i.e., with
probability equal to unity) the value of a physical quantity, then there exists an element of physical reality corresponding lo this physical quantity.}
\flushright
 (Einstein, Podolsky, and Rosen~1935)
\end{quote}
 It is true that Alice and Bob, after measuring and collecting the results of their  spin $x$ measurements, can predict with certainty the outcome of Charley's spin $x$ measurement. However,  believing in the MWI, Alice and Bob know that their prediction is not universally true. It is true only in their particular world. They {\it know} that there are  parallel worlds in which Charley's outcome is different. There is  no ``counterfactual definiteness'' which is frequently assumed in Bell-type arguments: no definite outcome   exists  prior to the measurement.

To see explicitly how the MWI removes the action at a distance of Bell-type experiments consider a demonstration of the GHZ experiment which is not yet possible with current technology, but may become possible in the near future. The choice of which components of the spin are measured  by Alice, Bob, and Charley  is made according to the ``random''  results of other quantum measurements \cite{Sheid}.   I argued that a better strategy is to rely  on macroscopic signals from galaxies in different parts of the Universe \cite{VBell}, but for my analysis here, a quantum device is more appropriate.

In the GHZ setup analysis not all combinations of measurements are considered: spin $x$ measurements are performed by just one  observer or by all three observers. To ensure this we distribute between Alice, Bob and Charley another GHZ set of  spin-$\frac{1}{2}$ particles. So, the state of all particles  (in the $z$ basis) is:
\begin{equation}
\label{GHZ+}
  {1\over
  2}{\Large (}|{\uparrow}\rangle_A|{\uparrow}\rangle_B|{\uparrow}\rangle_C -
|{\downarrow}\rangle_A|{\downarrow}\rangle_B|{\downarrow}\rangle_C{\Large
)}~{\Large (}|{\uparrow}\rangle_A|{\uparrow}\rangle_B|{\uparrow}\rangle_C -
|{\downarrow}\rangle_A|{\downarrow}\rangle_B|{\downarrow}\rangle_C{\Large
)} .
\end{equation}
Alice, Bob and Charley perform spin $x$ measurement of their additional spins. If the outcome is 1,  then the spin $y$ measurement of the second particle, the one from the original GHZ set, is performed. If the outcome is -1, then the   $x$ component of the spin is measured instead.

Alice's measurements split the world into four worlds. According to Everett's ``relative state formulation of quantum theory'' \cite{Eve}, these  are   Alice's worlds: $(\uparrow_{xA}),(\downarrow_{xA}),(\uparrow_{yA}),(\downarrow_{yA})$.   Nothing changes at Bob's and Charley's sites because of Alice's actions. The complete local descriptions of Bob's and Charley's GHZ spins remain to be  the same mixtures: completely unpolarized spins.

Now let us add the measurements of Bob. He also splits his world  into four Everett worlds. According to the EPR argument, after Alice's  and Bob's measurements, there is an element of reality associated with Charley's spin. Indeed, the information about the results of  their measurements tells us what is Charley's spin. For example, in the  world  $(\uparrow_{xA}, \uparrow_{xB})$,  Charley's spin is $|{\downarrow}\rangle_x$. But in a parallel world $(\uparrow_{xA}, \downarrow_{xB})$,  Charley's spin is $|{\uparrow}\rangle_x$. So, there is no single element or reality of Charley's spin $x$ in Nature.

 According to the definition of a ``world'' that I prefer \cite{SEP}, in any world all macroscopic objects  have well localized states.  All measuring devices show definite values, so in every world Alice, Bob and Charley have well defined values of spins. Alice first splits the world into four different worlds according to her measurements. Each of the worlds is then split again into four worlds by Bob. Charley, however, does not make any additional splitting. In every one of the 16 worlds created by Alice and Bob, the outcomes of his two spin measurements are already fixed. Here are all 16 worlds:
 \begin{eqnarray}
\nonumber
 (\downarrow_{xA}, \downarrow_{xB}, \downarrow_{xC}),~~(\downarrow_{xA}, \uparrow_{xB}, \uparrow_{xC}),~~(\uparrow_{xA}, \downarrow_{xB}, \uparrow_{xC}),~~(\uparrow_{xA}, \uparrow_{xB}, \downarrow_{xC}),~~ \\ \nonumber
 (\uparrow_{xA}, \uparrow_{yB}, \uparrow_{yC}),~~(\uparrow_{xA}, \downarrow_{yB}, \downarrow_{yC}),~~(\downarrow_{xA}, \uparrow_{yB}, \downarrow_{yC}),~~(\downarrow_{xA}, \downarrow_{yB}, \uparrow_{yC}),~~ \\
 (\uparrow_{yA}, \uparrow_{xB}, \uparrow_{yC}),~~(\uparrow_{yA}, \downarrow_{xB}, \downarrow_{yC}),~~(\downarrow_{yA}, \uparrow_{xB}, \downarrow_{yC}),~~(\downarrow_{yA}, \downarrow_{xB}, \uparrow_{yC}),~~ \\\nonumber
 (\uparrow_{yA}, \uparrow_{yB}, \uparrow_{xC}),~~(\uparrow_{yA}, \downarrow_{yB}, \downarrow_{xC}),~~(\downarrow_{yA}, \uparrow_{yB}, \downarrow_{xC}),~~\downarrow_{yA}, \downarrow_{yB}, \uparrow_{xC}).~~
  \end{eqnarray}

All the worlds  fulfill equations  (\ref{E1}-\ref{E4}). However, we do not get a contradiction as in Section \ref{E-B-G} because the equations do not have to be correct together: each one of the equations is correct in  four worlds for which it can be applied. Different equations are valid in different worlds, so that the  values of the spins in the equations can be different. The contradiction arises  if we assume that there is only one world.

\section{MWI and nonlocality }

As shown above, the MWI removes action at a distance from quantum physics. Like in classical relativistic physics,   any local action on a system changes nothing whatsoever at remote locations at the moment of disturbance. It does not mean, however, that quantum mechanics provides a local picture similar to classical physics with particles and fields localised in 3-space.

In classical physics, the complete description is given by specifying the trajectories of the particles and values of the fields:
\begin{equation}\label{ClUni}
{\rm Universe} = \{\vec{r}_i(t),~ \vec F_j(\vec{r},t)\}.
\end{equation}
It is local because it can be alternatively presented as an infinite set of vectors for all space-time points $(\vec{r},t)$ which provide values of projection operators $\{{\rm \bf P}_i(\vec{r},t)\}$ and values of all fields at this point $\{\vec F_j(\vec{r},t) \}$.

In classical physics, outcomes of an experiment at every site are fully specified by the local description of this site. The measuring devices and the observers can be expressed in the same language, in terms of the locations  $\vec{r}_i$s of the particles that they are made of. The final positions of the particles of the measuring devices are fully explained by their initial states and by the local interactions occurring  in the interval between the initial and final times.   Classical physics is deterministic (classical probability theory is relevant only for situations with incomplete knowledge of the full description), so the issue of correlations between outcomes of experiments at different places does not arise.

In a quantum world, if all particles are in a product state, then the description of the Universe is similar to the classical one: it is a set of these wave functions,
\begin{equation}\label{prodUni}
{\rm Universe}  = \{\Psi_i(\vec{r},t)\},
\end{equation}
which also can be represented as an infinite set of vectors with values of the wave functions at all space-time points. (For the current analysis it is not necessary to go to field theory which describes quantum fields). However, if we introduce measuring devices and observers, the local coupling of the measurement process will destroy the above product state. The description of the Universe then is the wave function in the configuration space of all particles,
\begin{equation}\label{Uni}
{\rm Universe}  =  \Psi(\vec{r}_1, \vec{r}_2,...\vec{r}_i,...,t) ,
\end{equation}
 which cannot be represented   as a set of vectors in  space-time points.

 In the MWI, this wave function in the configuration space is all that exists. It  explains everything, but not in a simple way. It does not provide a transparent connection to  our experiences. The way to connect the Universal wave function to our experience is to decompose it into a superposition of terms, each one corresponding to a different world. In each such term all    variables specifying states of macroscopic objects are essentially in a product state. 
 
 In each of the 16 worlds of our  GHZ experiment,  Alice, Bob and Charley have definite results of their spin measurements. What makes this situation nonlocal is that while all four different local options are present for all observers, i.e., there are four Everett worlds for Alice, and separately for  Bob and for Charley, we do not have 64 worlds. Specifying Everett worlds of two observers fixes the world of the third. This connection between local worlds of the observers is the nonlocality of the MWI.

 Is there any possibility for an action at a distance in the framework of the MWI? Obviously, on the level of the physical Universe which includes all the worlds, local action cannot change anything at remote locations. However, a local action splits the world which is a nonlocal concept, and local actions can make splitting to worlds which differ at remote locations. Thus, an observer for whom only his world is relevant,  has an illusion of an action at a distance when he performs a measurement on a system entangled with a remote system. (He also has an illusion of randomness each time he performs a quantum measurement \cite{qmdet}.)

 Consider again our example when Alice and Bob finished their measurements, but Charley still did not make his measurements. He knows that Alice and Bob made the measurements, he knows that there are 16 worlds:
 \begin{eqnarray}
\nonumber
 (\downarrow_{xA}, \downarrow_{xB}),~~(\downarrow_{xA}, \uparrow_{xB}),~~(\uparrow_{xA}, \downarrow_{xB}),~~\uparrow_{xA}, \uparrow_{xB}),\\ \nonumber
 (\uparrow_{xA}, \uparrow_{yB}),    ~~(\uparrow_{xA}, \downarrow_{yB}),
 ~~(\downarrow_{xA}, \uparrow_{yB}),~~(\downarrow_{xA}, \downarrow_{yB}), \\
 (\uparrow_{yA}, \uparrow_{xB}),~~(\uparrow_{yA}, \downarrow_{xB}), ~~(\downarrow_{yA}, \uparrow_{xB}),~~(\downarrow_{yA}, \downarrow_{xB}),\\
 (\uparrow_{yA}, \uparrow_{yB}),  ~~(\uparrow_{yA}, \downarrow_{yB}),~~(\downarrow_{yA}, \uparrow_{yB}),~~(\downarrow_{yA}, \downarrow_{yB}).\nonumber
  \end{eqnarray}
 Charley is in all these worlds. He is in a single Everett world which includes 16 worlds according to my definition. There is no meaning to ask him now in which world out of 16 he is.

 If Charley follows the instructions and performs the measurements according to the rules stated above, he will create four Everett worlds by creating macroscopic outcomes in his laboratory. Each of his new Everett worlds belongs to four worlds specified by Alice and Bob.
  
  Charley has a choice of performing or not performing the measurements and this will change the set of worlds he  will belong to. He can also make spin measurements not in accordance to the instructions: for example, he can make spin $y$ measurements instead of $x$ measurements and vice versa. Now, for every outcome he will end up belonging to eight, instead of four, worlds. He will also increase the total number of existing worlds to 32. If instead of following the instructions, all observers will make both spin measurements in the $z$ direction, there will be only 4 worlds:
\begin{eqnarray}
\nonumber
 (\downarrow_{zA},\downarrow_{zA},\downarrow_{zB},\downarrow_{zB}, \downarrow_{zC}, \downarrow_{zC}),~~
  (\downarrow_{zA},\uparrow_{zA},\downarrow_{zB},\uparrow_{zB}, \downarrow_{zC}, \uparrow_{zC}),~~\\
   (\uparrow_{zA},\downarrow_{zA},\uparrow_{zB},\downarrow_{zB}, \uparrow_{zC}, \downarrow_{zC}),~~
    (\uparrow_{zA},\uparrow_{zA},\uparrow_{zB},\uparrow_{zB}, \uparrow_{zC}, \uparrow_{zC}).~~
   \end{eqnarray}
Each observer, performing measurements in  $z$ direction creates worlds in which other observers have definite values of spin in $z$ direction. It looks like nonlocal action at a distance: local measurement changed some property in remote location. But it is a subjective change for an observer in a particular world: he  understands that in the physical universe which includes worlds with all outcomes of his local measurements, the remote spins also have all possible values.

\section{Conclusions}

Bell inequalities lead us to a hard choice: either we believe that there is some kind of action at a distance, or that there are multiple realities. My strong feeling is that accepting action at a distance is the bigger price and I am convinced that the MWI is the correct description of Nature.

In the MWI  the Bell proof of action at a distance fails in an obvious way since it requires a single world to ensure that measurements have single outcomes. Although there is no action at a distance in the MWI, it still has nonlocality.
The core of the nonlocality of the MWI is entanglement which is manifested in the connection between local Everett worlds of the observers. I feel that Bell inequalities can be manifested as a property of these connections, but I could not find a simple way to formulate it. I hope that this will be done in the future.

My first formulation of the MWI and arguments in its favor appeared in a preprint \cite{schizo} that I sent to John Bell at the end of 1989. He was not convinced. He replied with a short paragraph saying that if there are multiple worlds, there should be one in which I do not believe in the MWI. He added, more seriously, that he does not know what is the right way to understand quantum mechanics, but the MWI does not sound plausible to him. He expressed this view  in more details in the Nobel Symposium  \cite{Bell86}:
  \begin{quote}
 The `many world interpretation seems to me an extravagant, and above all an extravagantly vague, hypothesis. I could almost dismiss it as silly. And yet... It may have something distinctive to say in connection to `Einstein Podolsky Rosen puzzle', and it would be worthwhile, I think, to formulate some precise version of it to see if it really so. And the existence of all possible worlds may make us more comfortable about existence of our own world... which seems to be in some ways a highly improbable one.\flushright
(John Bell, 1986)
\end{quote}

For me Bell's result  was the first   reason to accept the MWI. Since then, the discovery of  teleportation and of the interaction-free measurements turned my belief into a strong conviction \cite{PSA}. I feel that now I developed ``the precise version of the MWI'' which John Bell was looking for \cite{qmdet}. I regret that I did not have this clear vision at 1989 when I discussed interpretation of quantum mechanics with John Bell in Erice.

I thank Eliahu Cohen  and Shmuel Nussinov for helpful discussions. This work has been supported in part by   the Israel Science Foundation  Grant No. 1311/14  and German-Israeli Foundation  Grant No. I-1275-303.14.

\end{document}